# Secured Message Transmission in Mobile AD HOC Networks through Identification and Removal of Byzantine Failures

[1]V. Anitha, [2]Dr. J. Akilandeswari
[1]Assistant Professor, Dept. of Electronics and Communication, Dayananda Sagar College of Engineering, Bangalore, India.
anithavijaya@gmail.com
[2]Professor, Dept. of Computer Science and Engineering, Sona College of Technology, Salem, India.
akila_rangabashyam@yahoo.com

**Abstract –** The emerging need for mobile ad hoc networks and secured data transmission phase is of crucial importance depending upon the environments like military. In this paper, a new way to improve the reliability of message transmission is presented. In the open collaborative MANET environment, any node can maliciously or selfishly disrupt and deny communication of other nodes. Dynamic changing topology makes it hard to determine the adversary nodes that affect the communication in MANET.

An SMT protocol provides a way to secure message transmission by dispersing the message among several paths with minimal redundancy. The multiple routes selected are known as APS –Active Path Set. This paper describes a technique for fault discovery process to identify Byzantine failures which include nodes that drop, modify, or mis-route packets in an attempt to disrupt the routing service. An adaptive probing technique detects a malicious link through binary search and according to the nodes behavior, these links are avoided in the active path by multiplicatively increasing their weights. The proposed scheme provides secure communication even with increased number of adversaries.

## 1. INTRODUCTION

Mobile Ad hoc networks are advantageous in situations where there are no network infrastructures available and when there is a need for people to communicate using mobile devices [5]. In mobile ad hoc networks, there is no central administration to take care of detection and prevention of anomalies. Therefore nodes have to cooperate for the integrity of the operation of the network. However, nodes may refuse to cooperate by not forwarding packets to others for selfish reasons and not want to exhaust their resources.

The SMT protocol [1] safeguards pair wise communication across an unknown frequently changing network, possibly in the presence of adversaries that may exhibit arbitrary behavior. The scheme presented in this paper guarantee that a Byzantine fault will be identified and the fault link can be avoided in the data transmission phase. The current topological information will be gathered based on the network behavior such as transmission time, probability of lost packets and acknowledged packets. Based on the information gathered, the route metrics are updated and used in selecting the multiple paths [4]. In this way the reliability of the secured data transmission can be enhanced.

## 2. PROBLEM DEFINITION

### 2.1 Data Transmission in MANET

In any type of MANETs, reliable delivery of information to the intended destination is of major interest to users sending information across that network. The information on the network might not be delivered to the destination as it is disrupted by the system because of many reasons. These reasons can be grouped into two categories, network faults and security attacks. In the former, main problem is to detect abnormal changes in the network and categorize them. Security attacks [19], [20] can be protected and authenticated by cryptography.

However, cryptographic protection cannot be effective against network layer attacks especially like Byzantine attacks.

A compromised intermediate node or a set of compromised intermediate nodes works in collusion and carries out attacks such as creating routing loops, routing packets on non optimal paths and selectively dropping packets are referred to as Byzantine attacks [10]. Byzantine failures are hard to detect. The network would seem to be operating normally in the viewpoint of nodes, though it may actually be exhibiting Byzantine behavior [6]. In contrast, the goal of a Byzantine node is to disrupt the communication of other nodes in the network, without regard to its own resource consumption [7].

Black Hole attack [18] is a basic Byzantine attack, where the adversary stops forwarding data packets, but still participates in the routing protocol correctly. As a result, whenever the adversarial node is selected as part of a path by the routing protocol, it prevents communication on that path from taking place. Most existing secure and insecure routing protocols are disrupted by black hole attacks because they render the normal methods of route maintenance futile.

In Byzantine Wormhole attack more than one node is compromised, it is reasonable to assume that these nodes may interact in order to gain an additional advantage. This allows the adversary to perform a more effective attack. Indeed, one such attack is a Byzantine wormhole, where two adversaries collude by tunneling packets between each other in order to create a wormhole in the network [21].

The adversaries can send a route request and discover a route across the ad hoc network, then tunnel packets through the non-adversarial nodes to execute the attack. The adversaries can use the low cost appearance of the wormhole links in order to increase the probability of being selected as part of the route, and then attempt to disrupt the network by dropping all of the data packets. Figure 1 shows the Byzantine wormhole attack, this is an extremely strong attack that can be performed even if only two nodes have been compromised.

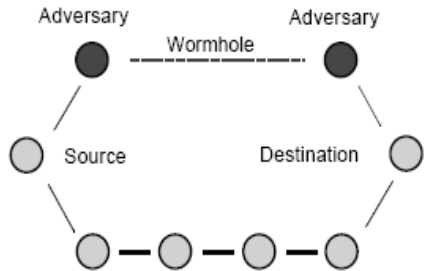

**Figure 1.** Byzantine Wormhole Attack

Byzantine wormhole attack considered in this work is different from the traditional wormhole attack. In the traditional wormhole attack, an adversary or multiple adversaries trick two honest nodes into believing that there exists a direct link between the honest nodes. The difference is that in the Byzantine case, the wormhole link exists between the adversarial nodes, not between the honest nodes.

## 3. RELATED WORK

The existing protocol SMT [1] for secured data communication provides end to end secure and robust feedback mechanism. A set of diverse and node disjoint paths are used at a time for data transmission and are known as Active Path Sets. The message and the redundancy are divided into a number of pieces, so that even a partial reception is able to reconstruct the data, called as Message dispersion [8] as shown in figure 2. The source updates the ratings of the paths based on the feedback.

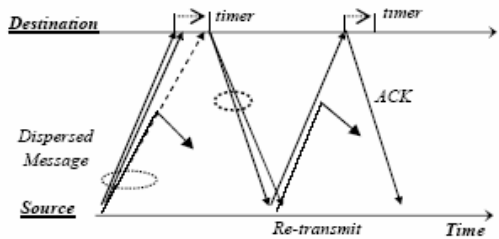

**Figure 2.** Message Dispersion in SMT

Secure Routing Protocol [11] is an overlay security layer for already existing routing protocol such as DSR [12]. The protocol bases its security on the assumption that the source and the destination share an authentication primitive. The routing messages are protected by nodes end-to-end authentication. SRP uses the route redundancy between the two end-hosts to increase the robustness against malicious nodes. The network is protected against flooding of route requests by limiting the rate of route request processed at each node. The problem is that the neighbors are not authenticated so a malicious node can blackmail other nodes by sending forged route requests. Besides, SRP does not make any assumption on the route request propagation algorithm as it is using the underlying protocol's one which is not optimized to find disjoint paths.

Although the selection of paths based on reliability as a routing metric has not been proposed before, other protocols have examined how to relate path availability to routing decisions. The (α,t)-Cluster algorithm [13] uses path availability to organize a MANET into clusters and provide a hybrid routing approach that is proactive within the cluster and reactive between clusters. While this protocol can group nodes into clusters among which path availability can be bounded, it does not help the selection among competing paths. Signal Stability-Based Adaptive (SSA) [14] and Associativity-Based Routing (ABR) [15] protocols propose two different mechanisms for assessing link stability. They both rely on periodic beaconing in order to estimate the link failure rates. SSA nodes collect statistics of the quality of their incident links based on the link utilization. Both protocols employ on-demand flooding of the network with route requests, which accumulate the corresponding stability metrics, and the destination chooses the most stable route. However, none of these protocols investigate the use of multiple paths or a way to select a set of such redundant paths.

## 4. THE PROPOSED SCHEME

### 4.1 Overview

The block diagram shown in Figure 3 is the Overall view of proposed system. The security of data transmission can be increased by selecting most secured routes in Active Path Set (APS). To improve the performance of the secured message transmission, most reliable paths can be selected and included in active path set –APS. Two mechanisms can be provided to select the most reliable paths:

Detect the misbehaving nodes and report such events. Maintain a traffic pattern reflecting the past behavior of other nodes.

The best routes can be selected which comprises of nodes that do not have history of avoiding forwarding packets along established routes. The path as a whole which includes the genuine nodes will be discarded in the SMT path selection mechanism [3]. In this proposed scheme instead of avoiding the whole path only the links that are incident on the adversary nodes will be excluded. Upon detection of a misbehaving node, a report is generated and nodes update the rating of the reported misbehaving nodes. The ratings of the nodes along a well behaved route are periodically incremented while reception of a misbehavior node dramatically decreases node rating. When a new route is required the source node calculates the average of the ratings of the nodes in each of the route replies and selects the route with highest metric. In this system the routes will be categorized according to attributes such as length and whether the routes include any additional trusted nodes other than the destination.

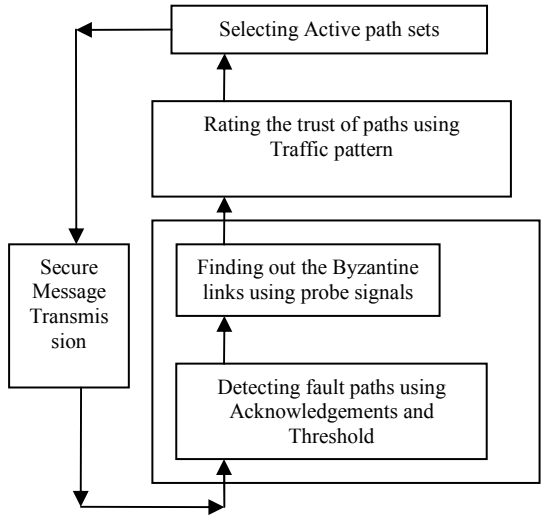

**Figure 3.** Overall view of proposed system

### 4.2 Route Discovery Process

Here both the route request and route discovery are flooded in the network. A digital signature [16] is used to authenticate the source. This signature will prevent unauthorized route request and avoids unnecessary flooding in the network. The route request is forwarded hop by hop and digital signatures are used at each hop to prevent an adversary from specifying an arbitrary path [2]. The route discovery phase consists of the following phases.

**Request Initiation**

The source creates and signs a request that includes the destination, the source, a sequence number and a weight list. The source then broadcasts this request to its neighbors. The source's signature allows the destination and intermediate nodes to authenticate the request and prevents false route requests.

**Request Propagation**

The route requests are propagated by the intermediate nodes. Each request is checked against the list maintained by it. If it is distinct then the source signature is validated, stored in the list and then re – broadcasted. If the entire packet is verified then the node appends its identifier to the end of the packet and broadcasts the modified request.

**Response Initiation and Reception**

Upon receiving the request, destination node verifies the authenticity. It signs in the response and then broadcasts it. The source also performs the same verification as done by the intermediate nodes and stores in the source route list.

### 4.3 Byzantine Fault Detection

**Detection Scheme**

Proposed system discovers faulty links on the path from the source to the destination. An adaptive probing technique identifies a faulty link after log n faults have occurred, where n is the length of the path. It requires the destination to return an acknowledgement to the source for every successfully received packet. The source keeps track of the number of recent losses. If the number of recent losses violates the acceptable threshold, the system will register a fault between the source and the destination. Then a binary search will start on the path in order to identify the faulty link.

The source controls the search by specifying a list of intermediate nodes on data packets. Each node in the list in addition to the destination must send an acknowledgement to the source. The list of nodes those have to send acknowledgements are known as probe nodes. Since the list of probed nodes is specified on legitimate traffic, an adversary is unable to drop traffic.

**Binary Search in adaptive Probes**

The list of probes defines a set of non-overlapping intervals that covers the whole path where each interval covers the sub path between the consecutive probes that forms its end points. When a fault is detected on an interval, the interval will be divided into two by inserting a new probe. This new probe is added to the list of probes appended to future packets. The process of subdivision continues until a fault is detected on the interval that corresponds to a single link. This result in finding log n faults where n is the length of the path. From Figure 4, h node is the adversary node.

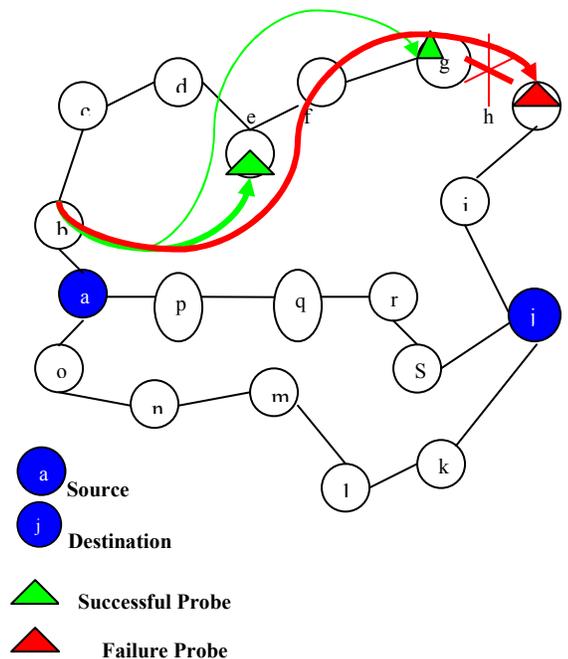

**Figure 4.** Binary Search- Finding Faulty Links Using Probe Signals

### 4.4 Key Management

Shared keys are used between the source and the probed nodes. This can be done by on demand Diffie-Hellmann key exchange algorithm [17]. This key mechanism is integrated into the route discovery protocol.

### 4.5 Calculating the Path Metric

After a sender and a receiver start to exchange data packets, they build tables to keep traffic patterns. Each table is composed of two fields: Packet identification number and time of action. Each time a packet is sent, the sender records the packet ID and the time. Each time a packet is received a receiver records the packet ID and the time.

Trip Time Variation (ΔTt)

Trip time of each packet is the time a packet spends on the way, starting when it is transmitted, ending when it is received. That time is calculated using the sender's time stamp when a packet was sent and recipient's time stamps when a packet was received.

$$Tt = Tr - Ts \qquad (4.1)$$

Here, Ts is the time a packet is sent and Tr the time it is received. In order to measure how long it would normally take a packet to travel from source to destination, estimated reference time variable (TE) is calculated. The estimated reference time can be given by the sum of the duration of the time of the route request message and the duration of the route reply message divided by two.

$$T_E = \frac{(R_{Q_r} - R_{Q_s}) + (R_{P_r} - R_{P_s})}{2} \qquad (4.2)$$

Having the TE value calculated in equation (4.2), we can calculate the Trip Time Variation (ΔTt) of each packet. The variation of trip time is measured with respect to the difference of estimated reference time and trip time experienced by the packet in the network.

$$\Delta Tt = TE - Tt \qquad (4.3)$$

Observing trip time variations over a period of time will allow the computation of probability of a packet to be delayed. Comparing trip time variation of many packets helps noticing and examining regular delays that are most likely to be caused by attacks

Change of packets frequency (ΔPf)

The sender compares both the frequency at which packets were sent and the frequency at which packets were received, measured in packets per second. By comparing the two frequencies, delays of packets can be noticed.

Lost Packets: (lp)

By looking at the anomaly detection table, it is easy to see packets that have been sent but not received. When there is a packet identification entry with sent time, but without received time, it shows that a packet was sent but not received. That packet could have been dropped after a time out on the queue, or it could have been dropped intentionally by an attacker. Anomaly Function is Af

$$Af = f(\Delta Tt, \Delta Pf, lp) \qquad (4.4)$$

Here ΔTt is trip time variation of packets, ΔPf is change of packet frequency, lp is lost packets.

**Link Failures**

Upon finding the link failure using binary search probes, all the paths containing that link will be discarded by increasing the link weights that are incident on those adversaries.

**4.6 Trust Updation and Path Set Selection**

An initial value is assigned to the variable of trust related to a path. A threshold is set to be based on expected behavior of the network environment. Based on the observation the paths metrics are updated and are used as a parameter while selecting the active path set.

## 5. EXPERIMENTAL RESULTS AND DISCUSSION

The experiments substantiate that the proposed system can, indeed successfully cope with a high number of adversaries, Active Path Set Secure Message Transmission (APS-SMT) can deliver many packets successfully than Non-Secure Protocol (NSP). APS-SMT is successful in delivering data with low, end-to-end delay.

Evaluation is done for two protocols (i) a single path data forwarding protocol that does not deploy any security mechanism to forward data called NSP (ii) In APS-SMT protocol, Multiple APS routes are discovered for SMT at the expense of increased overhead per route discovery, while single route discovered for NSP. OPNET simulation models were implemented.

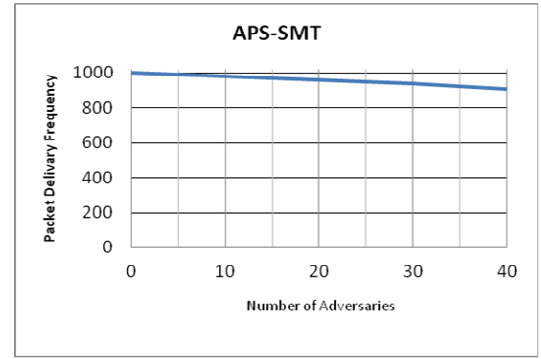

Figure 5a. Performance of APS-SMT

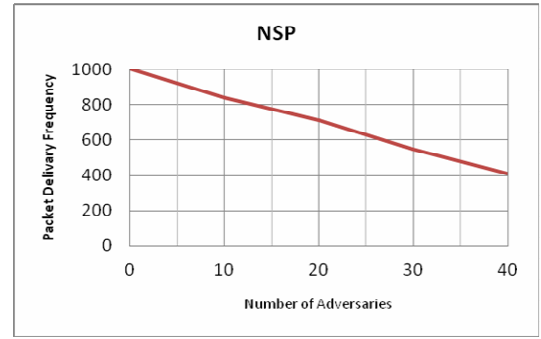

Figure 5b. Performance of NSP

Figure 5. Performance Evaluation of APS-SMT vs NSP

The network coverage is a 500m by 500m with 50 mobile nodes, with any two nodes able to communicate if they are within the reception distance which is set to 150m. The nodes are initially uniformly distributed throughout the network area and their movement is determined by the random way point mobility model [9].

The benefit of APS-SMT is clearly shown in the figure 5a. Thus APS-SMT delivers 99% of packets within the 5 to 15 adversaries and more than 95% of the packets even when 50% of the nodes are malicious. In contrast, the fast degradation of the NSP protocol as no surprise, as shown in figure 5b. The average APS-SMT improvement ranges from 32% to 150% as the number of adversaries increases.

## 6. CONCLUSION

In this paper, a design for an enhanced secured data transmission mechanism which is able to detect and remove the Byzantine failures has been presented. Path metric is evaluated by use of fault detection scheme, acknowledgements and trip time. Based on the evaluation, highly reliable paths can be chosen as multiple paths for SMT.

The successful delivery of message with the ability to disperse and avoidance of faulty links is more reliable than ordinary secured data transmission mechanism. It can be used in situations where reliability and security is most wanted like MANET in military.

**BIOGRAPHY**

**V.Anitha** has done her M.E and is Assistant Professor in the Department of Electronics and Communication, Dayanandasagar College of Engineering, Bangalore. She has a experience of 10 Years in Teaching and Research. She acquired her Bachelor's Degree in Electronics and Communication, from Kongu Engineering College, M.E Applied Electronics from Anna University. She has published papers in various International National Conferences. Her Areas of Interest includes MANETs, Network Security and Sensor Networks.

**Dr. J. Akilandeswari** is M.E. Ph.D. she is the Head of Department of Information Technology, Sona College of Technology, Salem. She has experience of 13 years in Teaching and Research. She is a Gold Medalist from Bharathidasan University University in Bachelor's of Computer Science and Engineering, M.E and Ph.D. in Computer Science and Engineering from NIIT, Tiruchirappalli. She has published papers in various international journals and Conferences. Her Areas of Interests includes Computer Networks, Distributed Computing, Data mining, and Web Mining.